\documentclass[twoside,slac_one]{revtex4}
\usepackage{graphicx}
\usepackage{fancyhdr}
\usepackage{amsmath}
\usepackage{bm}
\usepackage{amsxtra}
\usepackage{amssymb}
\usepackage{amsthm}
\usepackage{latexsym}
\usepackage{lscape}
\usepackage{subfigure}

\def\met{\ensuremath{E_{\mathrm{T}}^{\mathrm{miss}}}}
\def\Zprime{\ensuremath{Z^\prime}}
\def\pt{\ensuremath{p_{\mathrm{T}}}}

\def\et{\ensuremath{E_{\mathrm{T}}}}

\def\antibar#1{\ensuremath{#1\bar{#1}}}
\def\ttbar{\antibar{t}}
\def\ipb{\mbox{pb$^{-1}$}}

\begin{document}

\title{Search for New Physics Involving Top Quarks at ATLAS}

\author{T. Golling on behalf of ATLAS}
\affiliation{Department of Physics, Yale University, New Haven, CT, USA}

\begin{abstract}
  Two searches for new phenomena involving top quarks are presented: a
  search for a top partner in \ttbar\ events with large missing
  transverse momentum, and a search for \ttbar\ resonances in
  proton-proton collisions at a center-of-mass energy of 7~TeV. The
  measurements are based on 35~\ipb and 200~\ipb of data collected
  with the ATLAS detector at the LHC in 2010 and 2011, respectively.
  No evidence for a signal is observed. The first limits from the LHC
  are established on the mass of a top partner, excluding a mass of
  275 GeV for a neutral particle mass less than 50 GeV and a mass of
  300 GeV for a neutral particle mass less than 10 GeV. Using the
  reconstructed \ttbar{} mass spectrum, limits are set on the
  production cross-section times branching ratio to \ttbar{} for
  narrow and wide resonances.  For narrow \Zprime{} models, the
  observed 95\% C.L. limits range from approximately 38~pb to 3.2~pb
  for masses going from $m_{Z'}=$ 500 GeV to $m_{Z'}=$ 1300 GeV.  In
  Randall-Sundrum models, Kaluza-Klein gluons with masses below 650
  GeV are excluded at 95\% C.L.
\end{abstract}

\maketitle

\thispagestyle{fancy}

\section{Introduction}
The Standard Model of particle physics is believed to be an effective
theory valid up to energies close to 1 TeV. However, no new physics
beyond the Standard Model (SM) has been observed yet, and it is
critical to explore a wide range of possible signatures. A promising
avenue lies in final states that involve the heaviest of the particles
presumed to be elementary: the top quark. This document describes two
searches for new phenomena involving top quarks using the ATLAS
detector~\cite{AtlasDetector} at the CERN Large Hadron Collider (LHC).

The first search is carried out for a pair-produced exotic top partner
($T$), decaying to a top-antitop pair and two stable, neutral
weakly-interacting particles ($A_0$, which in some models may be its
own anti-particle)~\cite{ATLAS-CONF-2011-036}.  In most models, the
$T$ has typical quark-like quantum numbers, and is produced through
$q\overline{q}$ annihilation and gluon fusion. The final state for
such a process ($T\overline T \to t \overline t A_0 A_0$) is identical
to $t\overline{t}$, though with a larger amount of missing transverse
momentum (\met) from the undetected $A_0$'s. In supersymmetry models
with $R$-parity conservation $T$ refers to the stop squark and $A_0$
refers to the lightest supersymmetric particle. The
$t\overline{t}+$\met~\cite{Han:2008gy} signature appears in a general
set of dark matter-motivated models, as well as in other SM
extensions, such as the mentioned supersymmetry models, little Higgs
models with $T$-parity
conservation~\cite{Alwall:2010jc,ArkaniHamed:2001nc}, models of
universal extra dimensions (UED) with
Kaluza-Klein-parity~\cite{Appelquist:2000nn}, models in which baryon
and lepton number conservation arises from gauge
symmetries~\cite{FileviezPerez:2010gw} or models with third generation
scalar leptoquarks~\cite{Abazov:2008jp}. Many of these models provide
a mechanism for electroweak symmetry breaking and predict dark matter
candidates, which can be identified indirectly through their large
\met\ signature. The search is carried out in the $t\overline t$
lepton+jets channel where one $W$ boson from the top quark decay
decays leptonically (including $\tau$ decays to $e$ or $\mu$) and the
other $W$ boson decays hadronically, resulting in a final state with
an isolated lepton of high transverse momentum, four or more jets and
large \met.

A second search is carried out for new heavy particles decaying to
$t\overline t$ pairs~\cite{ATLAS-CONF-2011-087}. Using the lepton+jets
channel the reconstructed $t\overline t$ mass spectrum, based on three
or four jets, an electron or muon, and a neutrino, is used to search
for a signal.  Previous searches for $t\overline t$ were most recently
carried out by the CDF~\cite{:2007dz,:2007dia} and
D0~\cite{Abazov:2008ny} collaborations at Run II of the Fermilab
Tevatron Collider, and the ATLAS~\cite{ATLAS-CONF-2011-070} and
CMS~\cite{CMS-PAS-TOP-10-007} collaborations at the CERN LHC.  No
evidence for new particles was uncovered and 95\% confidence level
(C.L.) limits were set on the mass of a leptophobic topcolour $Z'$
boson~\cite{Hill:1993hs} at $m_{Z'} > 725$~\cite{:2007dz}
GeV\footnote{Preliminary results from CDF and D0 exclude a mass below
  900 and 820 GeV respectively.} as well as on the coupling strength
of a heavy colour-octet vector particle.  This analysis is very
similar to the previous ATLAS analysis~\cite{ATLAS-CONF-2011-070} but
uses data collected in 2011.  The benchmark model used to quantify the
experimental sensitivity to narrow resonances is a topcolour \Zprime{}
boson~\cite{Harris:1999ya} arising in models of strong electroweak
symmetry breaking through top quark condensation~\cite{Hill:1994hp}.
The specific model used is the leptophobic scenario, model {\bf IV} in
Ref.~\cite{Harris:1999ya} with $f_1=1$ and $f_2 =0$ and a width of
1.2\% of the $Z_t'$ boson mass.  This is the same set of parameters
that was used by the D0 Collaboration~\cite{Abazov:2008ny}.  The model
used for wide resonances is a Kaluza-Klein gluon $g_{KK}$, which
appears in Randall-Sundrum (RS) models with a warped extra dimension
in which particles are located in the extra
dimension~\cite{Lillie:2007yh,Djouadi:2007eg}.  The concrete model
used is described in detail in Ref.~\cite{ATL-PHYS-PUB-2010-008} and
implemented in the {\sc Madgraph}~\cite{Alwall:2007st} event
generator.  The couplings to quarks take the ``standard'' RS
values~\cite{Lillie:2007yh}: $g_L=g_R=-0.2$ for light quarks including
charm, $g_L=1.0, g_R=-0.2$ for bottom quarks and $g_L=1.0, g_R=4.0$
for the top quark.  In this case, the resonance is predicted to be
significantly wider than the detector and reconstruction algorithm's
resolution.  This model is taken as a proxy for coloured resonances.

\section{The ATLAS Detector}
The ATLAS detector~\cite{AtlasDetector} consists of an inner detector
tracking system (ID) surrounded by a superconducting solenoid
providing a $2$~T magnetic field, electromagnetic and hadronic
calorimeters, and a muon spectrometer (MS).  The ID consists of pixel
and silicon microstrip detectors inside a transition radiation tracker
(TRT) which provide tracking in the region $|\eta| < 2.5$.  The
electromagnetic calorimeter is a lead/liquid-argon (LAr) detector in
the barrel ($|\eta| < 1.475$)~\footnote{The azimuthal angle $\phi$ is
  measured around the beam axis and the polar angle $\theta$ is the
  angle from the beam axis. The pseudorapidity is defined as $\eta
  \equiv -\ln \tan(\theta/2)$.} and the endcap ($1.375 < |\eta| <
3.2$) regions.  Hadron calorimetry is based on two different detector
technologies. The barrel ($|\eta| < 0.8$) and extended barrel ($0.8 <
|\eta| < 1.7$) calorimeters are composed of scintillator/steel, while
the hadronic endcap calorimeters ($1.5 < |\eta| < 3.2$) are
LAr/copper.  The forward calorimeters ($3.1 < |\eta| < 4.9$) are
instrumented with LAr/copper and LAr/tungsten, providing
electromagnetic and hadronic energy measurements, respectively.  The
MS consists of three large superconducting toroids with 24 coils, a
system of trigger chambers, and precision tracking chambers which
provide muon momentum measurements out to $|\eta|$ of 2.7.

\section{Data Samples}
The search for a pair-produced exotic top partner is based on data
recorded by the ATLAS detector in 2010 using 35~pb$^{-1}$ of
integrated luminosity. The search for $t\overline t$ resonances is
based on data recorded in 2011 using 200~pb$^{-1}$.  The data were
collected using electron and muon triggers. Requirements of good beam
conditions, detector performance and data quality are imposed.

\section{Simulated Samples}
Monte Carlo (MC) event samples with full ATLAS detector
simulation~\cite{AtlasSimulation} based on the {\sc Geant4}
program~\cite{Geant4} and corrected for all known detector effects are
used to model the signal process and most of the backgrounds. The QCD
multi-jet background is modeled using data control samples rather than
the simulation. The $t\overline{t}$ and single top samples are
produced with {\sc MC@NLO}~\cite{MC_at_NLO}, while the $W$+jets and
$Z$+jets samples are generated with {\sc Alpgen}~\cite{ALPGEN}. {\sc
  Herwig}~\cite{Herwig} is used to simulate the parton shower and
fragmentation, and {\sc Jimmy}~\cite{Jimmy} is used for the underlying
event simulation. The diboson background is simulated using {\sc
  Herwig}. The inclusive $W$+jets and $Z$+jets cross sections are
normalized to NNLO predictions~\cite{FEWZ}, and the cross sections of
the other backgrounds are normalized to NLO predictions~\cite{MCFM}.

{\sc MadGraph}~\cite{Alwall:2007st} is used to simulate the signal process,
and {\sc Pythia}~\cite{Pythia} is used to simulate the parton shower
and fragmentation.  A grid of $T$ and $A_0$ masses is generated with
250~GeV $\le m(T) \le$ 350~GeV and 10~GeV $\le m(A_0) \le$ 100~GeV. In
this search, only on-shell tops are considered, and therefore grid
points where the mass difference between the $T$ and $A_0$ approaches
the top quark mass are excluded.  Each sample is normalized to the
cross section calculated at approximate NNLO in QCD using {\sc
  Hathor}~\cite{Hathor}, ranging from 20.5 pb for a $T$ mass of
250~GeV to 3.0 pb for a $T$ mass of 350~GeV.

Signal samples for \Zprime{} bosons decaying to $t\bar{t}$ are
generated using {\sc Pythia} allowing all top quark decay modes.
Samples of RS gluons are generated with {\sc Madgraph}, and showered
with {\sc Pythia}.

\section{Common Event Selection}
Electron and muon candidates are selected as for other recent top
quark studies using the lepton+jets signature~\cite{Aad:2010ey}.  Jets
are reconstructed using the anti-$k_t$~\cite{Cacciari:2008gp}
algorithm with a distance parameter $R$, defined as $\Delta R\equiv
\sqrt{\Delta\eta^2 + \Delta\phi^2}$, of 0.4. To take into account the
differences in calorimeter response to electrons and hadrons, a $\pt$-
and $\eta$-dependent factor, derived from simulated events and
validated with data, is applied to each jet to provide an average
energy scale correction~\cite{JESv16} back to particle level.  In the
calorimeter, the energy deposited by particles is reconstructed in
three-dimensional clusters. The energy of these clusters is summed
vectorially, and the projection of this sum in the transverse plane
corresponds to the negative of the \met~\cite{MET}.  Corrections to
the hadronic and electromagnetic energy scales, dead material and
out-of-cluster energy are applied, and in the case of reconstructed
muons, an additional correction is included.

Events are selected with exactly one electron or muon, where each
lepton is required to pass the following selection criteria. Electrons
are required to satisfy $\et > 20$~GeV ($\et > 25$~GeV for the
resonance search due to tighter trigger requirements in 2011) and
$|\eta| < 2.47$. Electrons in the transition region between the barrel
and the endcap electromagnetic calorimeters ($1.37 < |\eta| < 1.52$)
are removed. Muon candidates are required to satisfy $\pt > 20$~GeV
and $|\eta| < 2.5$.  Events with four or more reconstructed jets with
$\pt > 20$~GeV ($\pt > 25$~GeV for the resonance search) and $|\eta| <
2.5$ are selected.

\subsection{Additional Selection for Top Partner Search}
To reduce the $W$+jets background, events are required to have \met $>
80$~GeV and $m_{\mathrm{T}}$ $> 120$~GeV, where $m_{\mathrm{T}}$ is
the transverse mass of the lepton and missing energy~\footnote{The
  transverse mass is defined by the formula $ m_{\mathrm{T}} =
  \sqrt{2p^{\ell}_T \met (1 - \cos(\phi^{\ell} - \phi^{\met})) }$,
  where $p^{\ell}_T$ is the \pt\ (\et) of the muon (electron) and
  $\phi^{\ell}$ ($\phi^{\met}$) is the azimuthal angle of the lepton
  (\met).}.  Events with either a second lepton candidate satisfying
looser selection criteria or an isolated track with $\pt>12$~GeV are
rejected in order to reduce the contribution from $t\overline{t}$
dilepton events. In particular the isolated track veto is useful for
eliminating single-prong hadronic $\tau$ decays in $t\overline{t}$
dilepton events.

\subsection{Additional Selection for Resonance Search}
In the electron channel, \met{} must be larger than 35 GeV and
$m_\textrm{T}(\textrm{lepton},\met) > 25$ GeV. In the muon channel,
\met{} $> 20$ GeV and \met{}+$m_\textrm{T}(\textrm{lepton},\met) > 60$
GeV is required.  At least one of the selected jets must be tagged as
a $b$-jet.

Jets originating from $b$-quarks are selected by exploiting the long
lifetimes of $B$-hadrons (about 1.5 ps) leading to typical flight
paths of a few millimeters, which are observable in the detector. The
SV0 $b$-tagging algorithm~\cite{ATLAS-CONF-2010-099} used in this
analysis explicitly reconstructs a displaced vertex from the decay
products of the long-lived $B$-hadron.  Two-track vertices at a radius
consistent with the radius of one of the three pixel detector layers
are removed, as these vertices likely originate from material
interactions. A jet is considered $b$-tagged if it contains a
secondary vertex, reconstructed with the SV0 tagging algorithm, with
$L/\sigma(L) > 5.85$, where $L$ is the decay length and $\sigma(L)$
its uncertainty. This operating point yields a 50\% $b$-tagging
efficiency in simulated \ttbar\ events. The sign of $L/\sigma(L)$ is
given by the sign of the projection of the decay length vector on the
jet axis.

\section{Background Estimate for Top Partner Search}\label{BG1}
A summary of the background estimates and a comparison with the observed
number of selected events passing all selection criteria is shown in 
Table~\ref{t:bg}.  A total yield of 17.2 $\pm$ 2.6 events is expected from
SM sources, and 17 events are observed in data.  The background composition 
is similar in the electron and muon channels.

\begin{table}[ht!]
\begin{center}
  \caption[Background Summary]{Summary of expected SM yields including
    statistical and systematic uncertainties compared with the
    observed number of events in the signal region.  }
\begin{tabular}{r||c|c|c|c|c|c||c|c}
\hline\hline
Source & single lepton $t\overline{t}$/$W$+jets & dilepton $t\overline{t}$ & multi-jet & $Z$+jets &dibosons & single top & total & data \\ \hline
Number of events & 8.4 $\pm$ 1.6 & 7.6 $\pm$ 2.0 & 0.2 $\pm$ 0.6 & 0.4 $\pm$ 0.1 & 0.2 $\pm$ $<$0.1 & 0.4 $\pm$ 0.1 & 17.2 $\pm$ 2.6 & 17\\ 
\hline\hline
\end{tabular}\label{t:bg}
\end{center}
\end{table}

The dominant background arises from $t\overline{t}$ dilepton final
states, in which one of the leptons is not reconstructed, is outside
the detector acceptance, or is a $\tau$ lepton.  In all such cases,
the $t\overline{t}$ decay products include two high-\pt\ neutrinos,
resulting in large \met\ and $m_{\mathrm{T}}$ tails.  In MC, the
second lepton veto removes 45\% of the dilepton $t\overline{t}$ and
10\% of the single-lepton $t\overline{t}$ in the signal region.  The
veto performance is validated in the data in several control regions
both enhanced and depleted in dilepton $t\overline{t}$, and in all
cases the veto efficiencies in MC and data agree within 10\%.

The next largest backgrounds come from single-lepton sources,
including both $W$+jets and $t\overline{t}$ with one leptonic $W$
decay.  Both the normalization and the shape of the $m_{\mathrm{T}}$
distribution for this combined background are extracted from the data.
First, the yield of the single lepton background estimated from
simulation is normalized in the control region 60~GeV $<$
$m_{\mathrm{T}}$ $<$ 90~GeV to the data.  Next, the shape of the
$m_{\mathrm{T}}$ distribution in MC is compared with data in various
control regions, where events satisfy the signal event selection but
have fewer than four jets.  Further, events with identified $b$-jets,
based on lifetime $b$-tagging~\cite{Aad:2010ey}, are rejected in order
to reduce the dilepton \ttbar\ background, such that the samples are
dominated by $W$+jets events; the corresponding loss of single-lepton
\ttbar\ from this $b$-jet veto is accounted for in the systematic
uncertainties. Good agreement is observed between data and MC and
based on this agreement an uncertainty of 15\% is assigned on this
background.

In general, QCD multi-jet events do not produce large \met\ and
therefore fail the kinematic requirements for the signal
region. Although the QCD multi-jet background is expected to be small,
it is difficult to model with simulation. Therefore, data-driven
techniques similar to those described in~\cite{Aad:2010ey} are used to
estimate this background.  In both lepton channels the contribution to
the signal region is consistent with zero.

Other electroweak processes, single top, diboson production ($WW$,
$WZ$, and $ZZ$), and $Z$+jets are estimated using MC simulation,
normalized to the theoretical cross section and total integrated
luminosity. They have small production cross sections compared to
$t\overline{t}$ and $W$+jets, and are further suppressed by the
multiple-jet, \met\ and $m_{\mathrm{T}}$ selection criteria.

\section{Background Estimate for Resonance Search}
The following data-driven relative scale factors are applied to the
simulated $W$+light parton multiplicity samples: 0-parton exclusive
sample: 0.978 $\pm$ 0.004; 1-parton exclusive sample: 1.107 $\pm$
0.015; 2-parton exclusive sample: 1.147 $\pm$ 0.047; 3-parton
exclusive sample: 0.86 $\pm$ 0.16; 4-parton exclusive sample: 1.63
$\pm$ 0.44; 5-parton inclusive sample: 0.95 $\pm$ 0.58. These are
determined by fitting the observed jet multiplicity distribution in a
$W$+jets-dominated data sample selected by: exactly one lepton with
\pt~$>$~20 GeV, veto on the presence of any other lepton with
\pt~$>$~10~GeV, 30 $<$ \met\ $<$ 80 GeV, 40 $< M_{\mathrm T} <$ 80
GeV, and rejection of events in which at least one of the four hardest
jets is $b$-tagged.  The latter cut ensures that the sample is
orthogonal to the signal selection.  The sample is expected to contain
over 95\% $W$+jets events, and remaining contributions from other
Standard Model processes are subtracted using MC predictions prior to
the fit.  Note that the resulting scale factors are heavily
anticorrelated, which is taken into account in the evaluation of
systematic uncertainties.  The uncertainties in the scale factors
arise from the limited size of the data sample. This background's
normalization uncertainty in the signal sample is 35\%, driven by the
uncertainty in the relative event tagging probabilities between events
with two and four or more jets.

The QCD multi-jet background is estimated in the same way as described in
Section~\ref{BG1}. The \ttbar\ and the other electroweak backgrounds
are estimated using MC simulation, normalized to the theoretical cross
section and total integrated luminosity.

\section{Top Partner Search}
\begin{figure}
  \centering
  \subfigure[Lepton + 2 jets]{
    \includegraphics[width=0.47\textwidth]{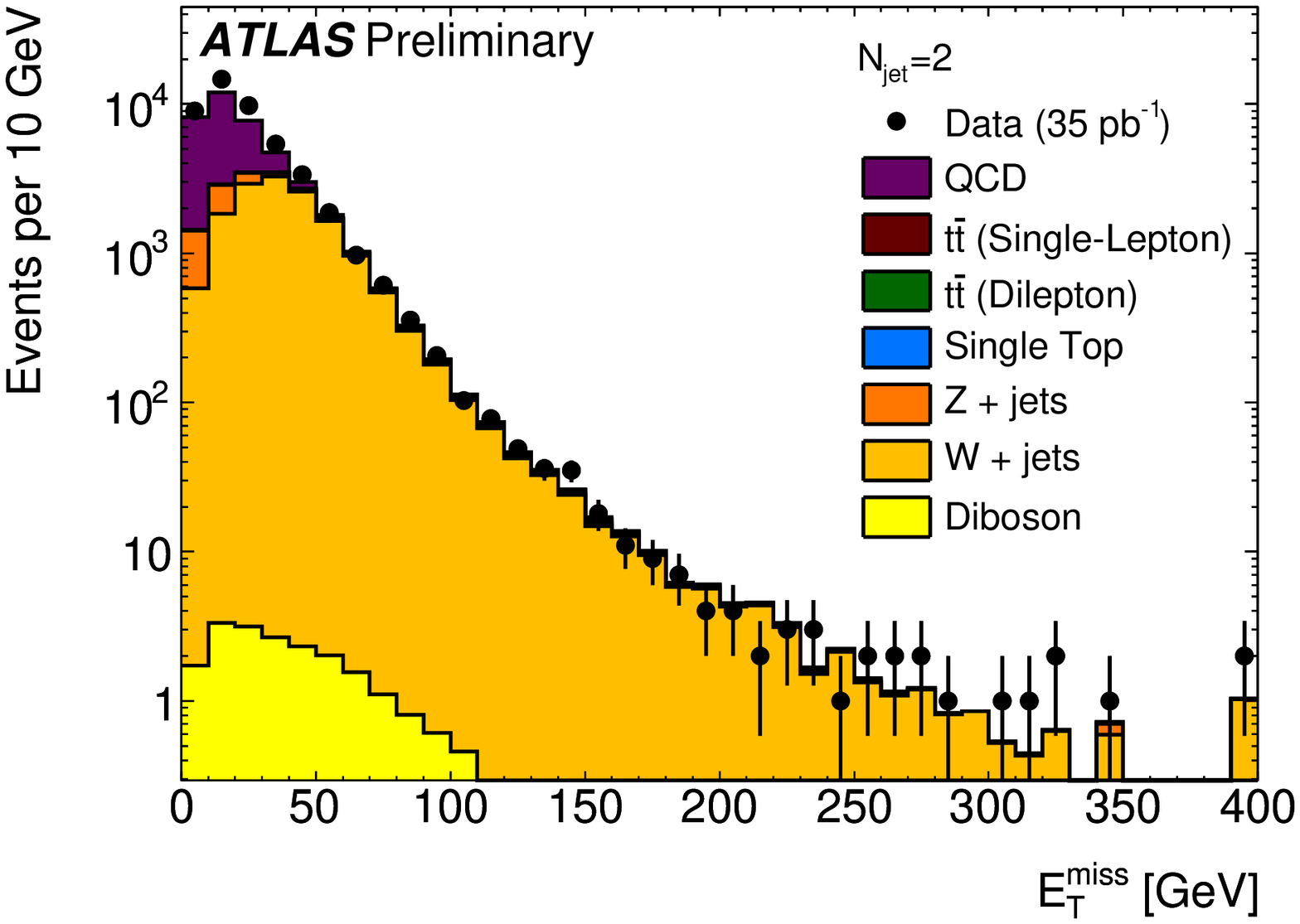}
  }
  \subfigure[Lepton + 3 jets]{
    \includegraphics[width=0.47\textwidth]{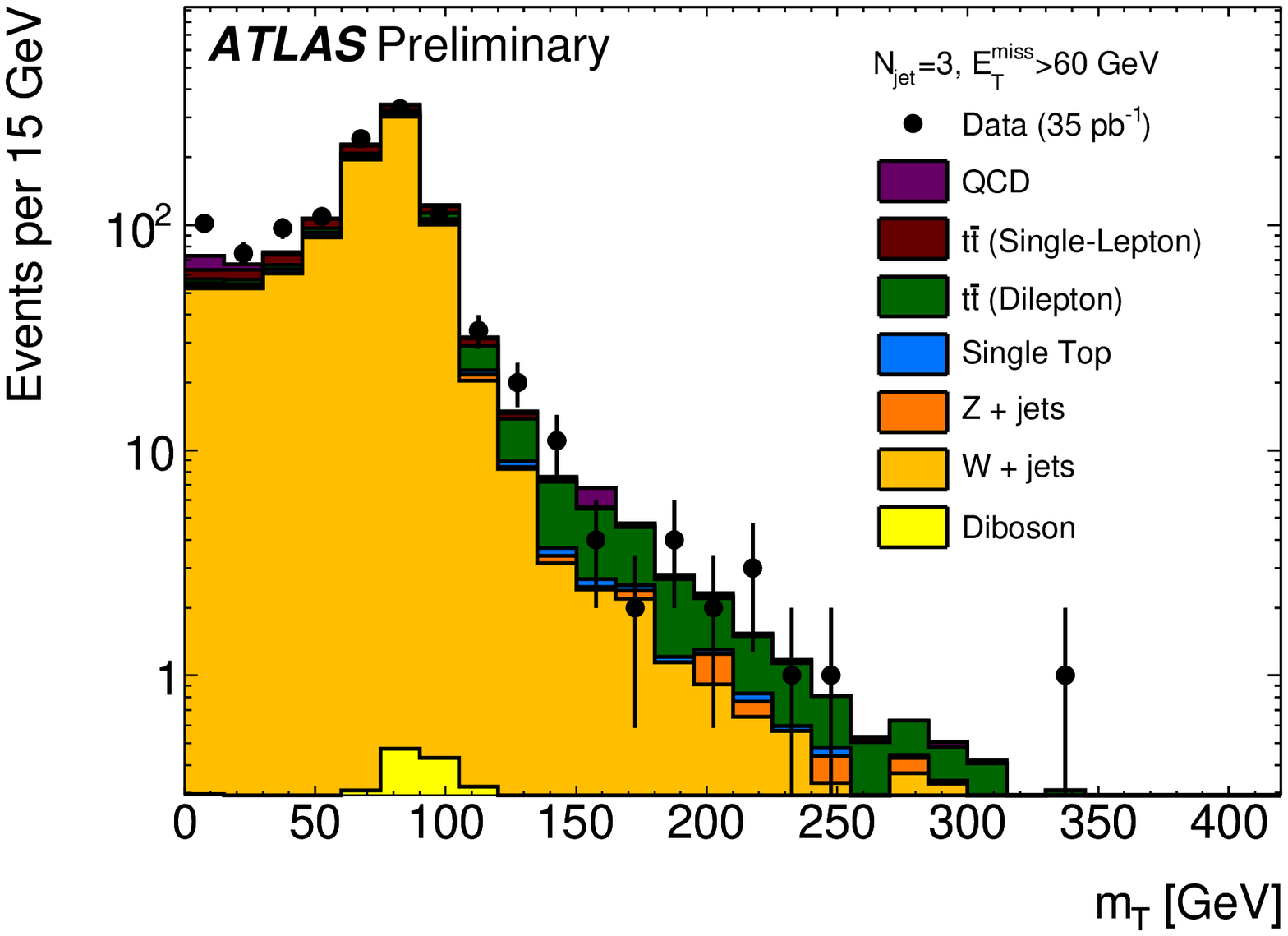}
  }
  \caption[$m_{\rm T}$, \met\ in Control Regions]{\met\ and $m_{\rm
      T}$ distributions in data and stacked simulation in two control
    samples of the top partner search, 2-jet events (a) and 3-jet
    events (b). \met $>$ 60 GeV is required in (b).}
\label{f:basickin}
\end{figure}

Due to the small size of the data sample, and in order to preserve as
much model-independence as possible, a simple cut-and-count analysis
is carried out in the high tails of the $W$ transverse mass and \met\
distributions. Figure~\ref{f:basickin} shows that the $W$ transverse
mass and \met\ tails are well modeled by SM contributions in two
different control samples.

\subsection{Systematic Uncertainties}
The dilepton veto carries an uncertainty of 15\%, determined in
control regions in data described in Section~\ref{BG1}. The single
lepton shape correction, $S(m_{\rm T})$, was determined to be
consistent with unity, but carries an uncertainty of 15\% from the
spread in control regions. The single lepton normalization is taken
from the $W$ mass range (60-90 GeV) just outside the signal region.
This normalization has an uncertainty of 10\%, which is mostly the
statistical uncertainty, but also includes the effect of varying the
mass range used. The jet energy scale and resolution have some
uncertainty~\cite{JESv16}, which is propagated through the analysis by
correcting the jet multiplicity, $m_{\rm T}$, and \met.  The per-jet
uncertainty ranges from more than 10\% at low $p_{\rm T}$ to roughly
5\% at high $p_{\rm T}$. The Monte Carlo is adjusted for measured
differences in lepton ID and trigger efficiencies, and these
corrections each contribute some uncertainty, typically 4-5\% per
lepton. The integrated luminosity has a relative uncertainty of 3.4\%.

\subsection{Results}
\begin{figure}
  \centering
  \subfigure[$m_{\rm T}$]{
    \includegraphics[width=0.47\textwidth]{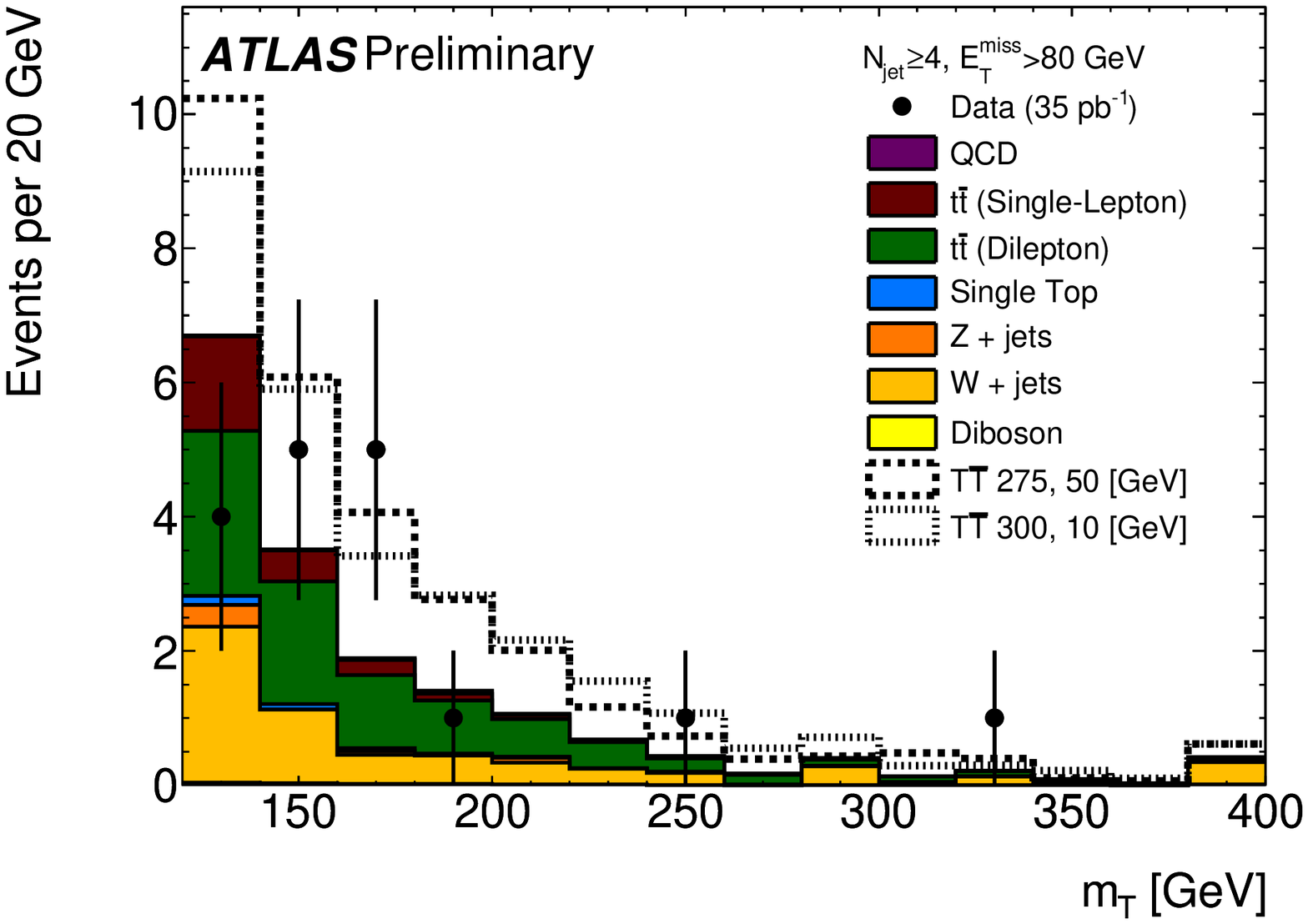}
  }
  \subfigure[\met]{
    \includegraphics[width=0.47\textwidth]{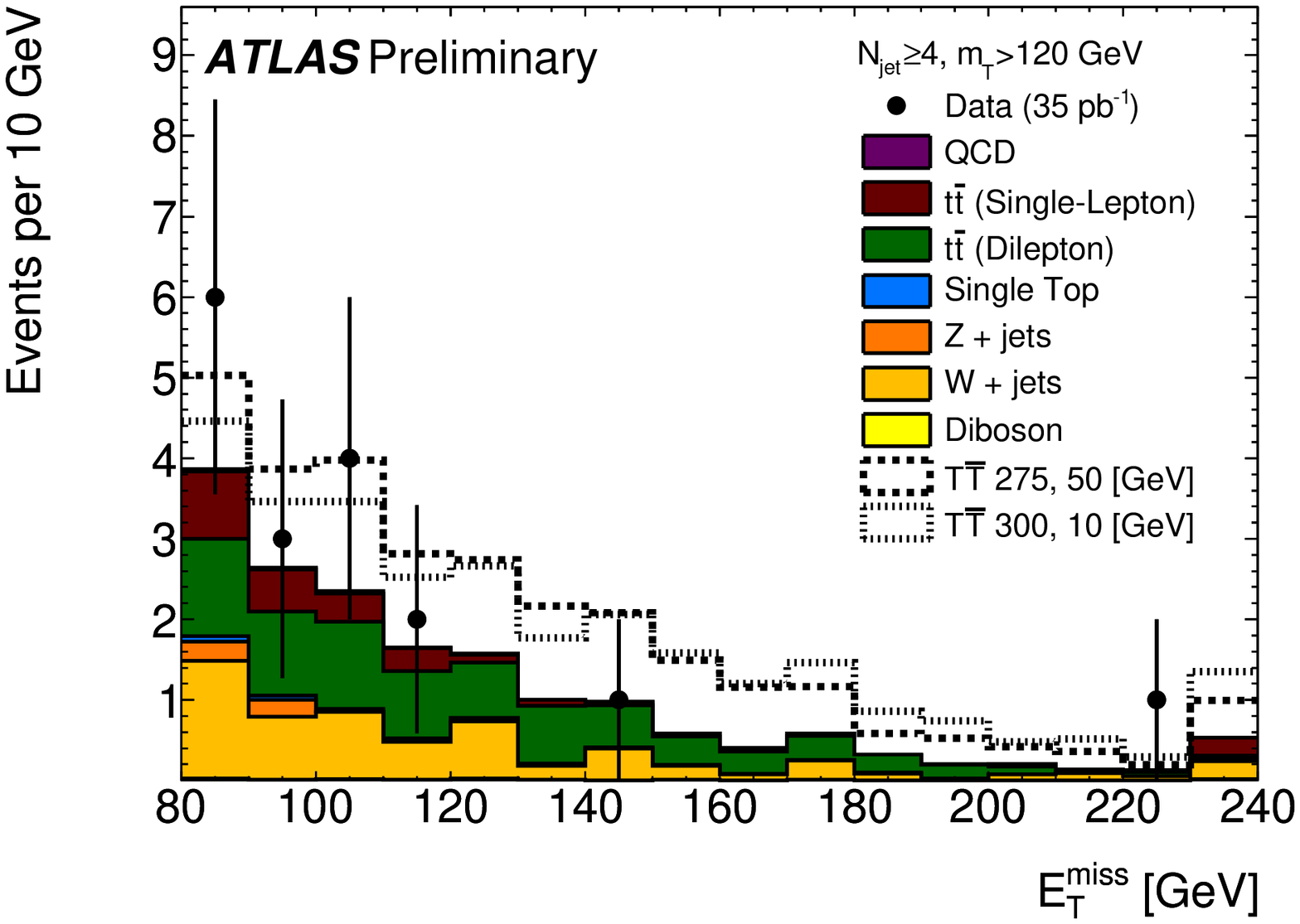}
  }
  \caption[Kinematics of Final Signal Sample in Data]{$m_{\rm T}$ and
    \met\ in the signal region of the top partner search.  The dotted
    and dashed lines show the expected distributions for two of the
    excluded signal mass points, where the numbers in the legends
    correspond to $m(T)$ and $m(A_0)$.}
\label{f:dataplots}
\end{figure}

Comparing with the data, good agreement with the combined background
prediction is observed: 17.2 background events are expected, and 17
events are observed. Figure~\ref{f:dataplots} shows some distributions
in the signal region.  No evidence of an excess or of a mis-modeled
background is observed. Given a theoretical cross section of 12 pb, an
additional 12.4 $\pm$ 3.1 signal events are expected from the 275 GeV
$T$ (50 GeV $A_0$) mass point.  Similarly, the 300 GeV $T$ (10 GeV
$A_0$) model would predict 11.7 $\pm$ 3.0 extra events from a cross
section of 7.3 pb~\cite{Berger}.  The event yield distribution is
studied from pseudo-experiments, assuming Gaussian systematics and
including correlations in signal and background systematics ({\it
  e.g.}, luminosity, theoretical cross sections), for both the signal
and background-only hypotheses, and it is determined that both of
these models can be excluded with confidence greater than 95\%.  The
samples with $m(T)$ ($m(A_0)$) of 250 (10) GeV and 275 (10) GeV are
excluded as well.  Implicitly, all other mass points with a lighter
$A_0$ are also excluded.

\section{\ttbar\ Resonance Search}

\subsection{Mass Reconstruction}
\begin{figure}[ht]
\centering
\subfigure[]
{
\includegraphics[width=0.47\textwidth]{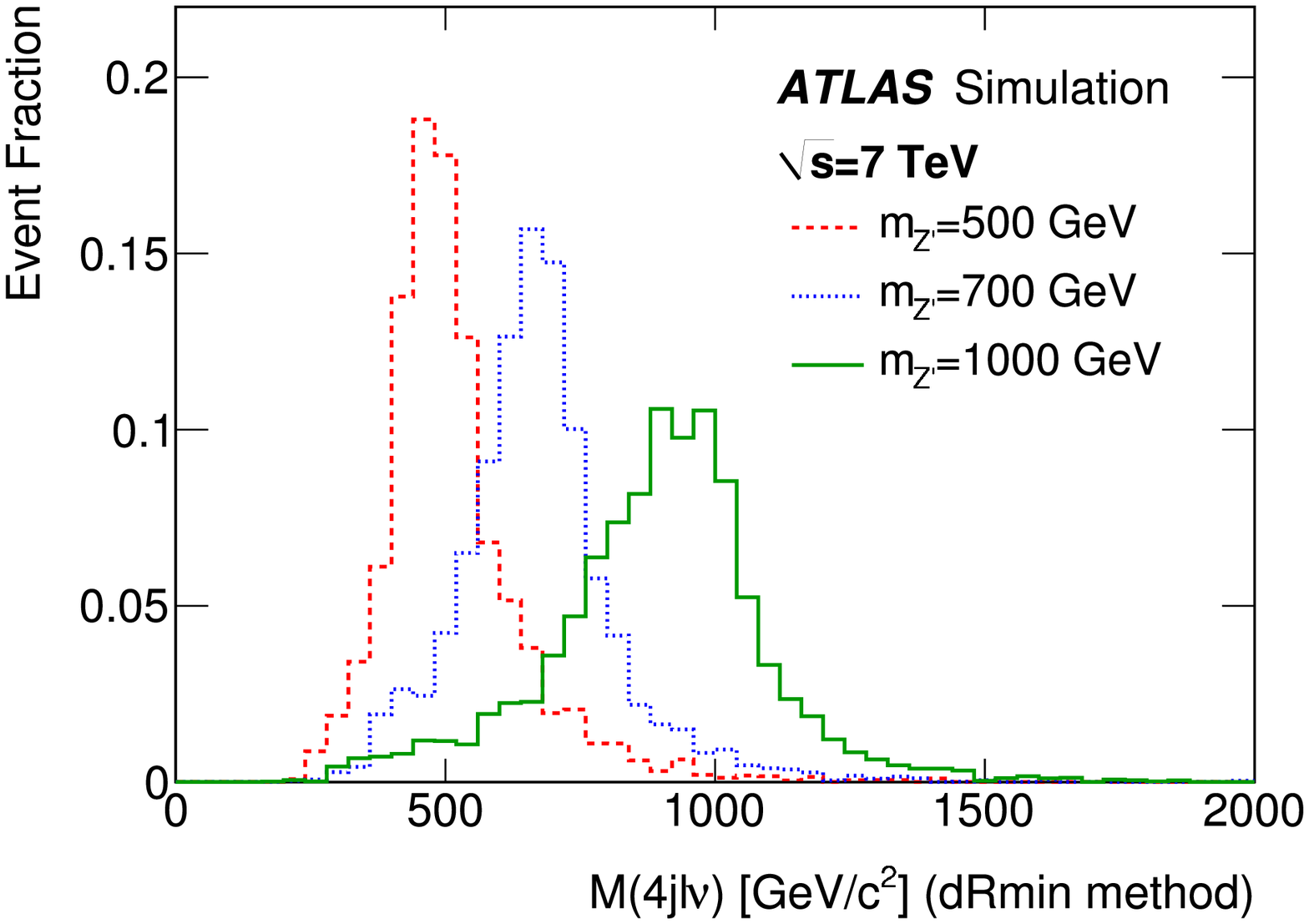}
}
\subfigure[]
{
\includegraphics[width=0.47\textwidth]{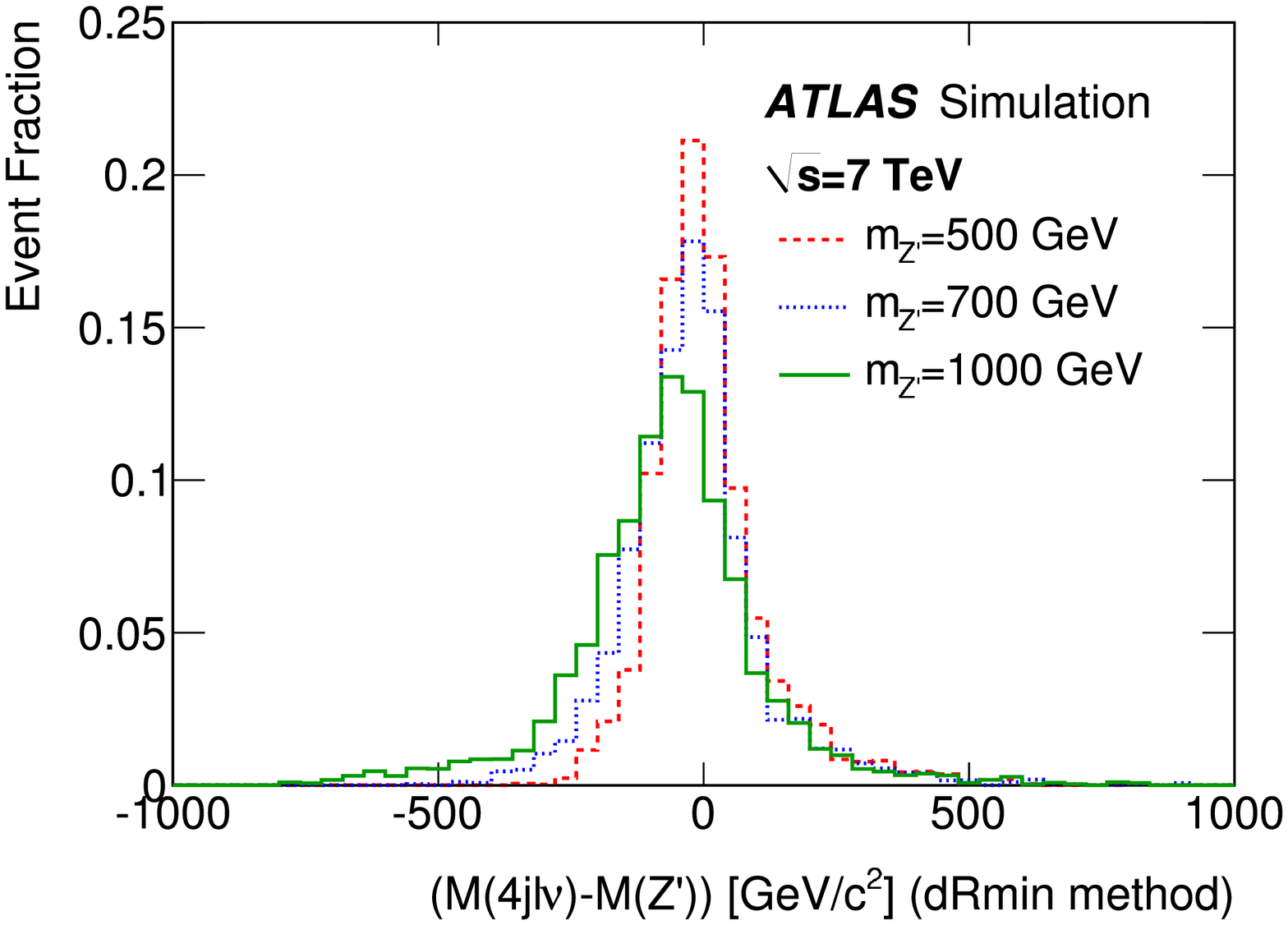}
}
\caption{Reconstructed $t\bar{t}$ pair invariant mass (a) and its
  resolution (b) for three $Z'$ boson masses: $m_{Z'} =$ 500 GeV,
  $m_{Z'} =$ 700 GeV and $m_{Z'} =$ 1000 GeV for the $dRmin$
  algorithms.}
\label{fig:drres}
\end{figure}

To reconstruct the $t\bar{t}$ mass, the neutrino's longitudinal
momentum ($p_z$) is determined by imposing the $W$-boson mass
constraint. If the discriminant of the quadratic equation is negative,
the missing transverse energy is adjusted to get a null
discriminant~\cite{chwalekphd}.  If there are two solutions, the
smallest $p_z$ solution is chosen.  The dominant source of long,
non-Gaussian tails in the mass resolution is the use of a jet from
initial- or final-state radiation in the place of one of the jets
directly related to a top quark decay product.  To reduce this
contribution, the $dRmin$ algorithm~\cite{ATLAS-CONF-2011-070}
considers the four leading jets with \pt\ $>$ 20 GeV and $|\eta| <$
2.5, and excludes a jet if its angular distance to the lepton or
closest jet satisfies $\Delta R_{min} > 2.5 - 0.015 \times m_j$, where
$m_j$ is the jet's mass.  (If more than one jet satisfies this
condition, the jet with the largest $dRmin$ is excluded.)  If a jet
was discarded and more than three jets remain, the procedure is
iterated.  Then $m_{t\bar{t}}$ is reconstructed from the lepton, \met\
and the leading four jets, or three jets if only three remain.  The
$\Delta R_{min}$ cut removes jets that are ``far'' from the rest of
the activity in the event.  Furthermore, by only requiring three jets
in the mass reconstruction, the method allows one of the jets from top
quark decay to be outside the detector acceptance, or merged with
another jet.  The reconstructed invariant masses and corresponding
resolutions obtained with the $dRmin$ algorithm are shown for three
different simulated $Z'$ boson masses in Fig.~\ref{fig:drres}.

\subsection{Systematic Uncertainties}
\label{sec:sys}
Systematic uncertainties that only affect the normalization of the
different backgrounds come from the uncertainty on the integrated
luminosity (4.5\%), background normalizations ($t\bar{t}$:
$^{+7.0}_{-9.6}$\%~\cite{Moch:2008qy}, single top: 10\%, $W$+jets:
35\%, diboson: 5\%, QCD multi-jet(e): 30\%, QCD multi-jet($\mu$):
50\%), and lepton trigger and reconstruction efficiencies ($\leq$
1.5\%).  The dominant shape uncertainties arise from the $b$-tagging
efficiency (11\% variation in the event yields), jet energy scale
including pileup effects (9\%)~\cite{JESv16}, and modelling of initial
and final state radiation (7\%).  The first two have been determined
from data by comparing results from different methods and/or data
samples, while the latter has been estimated from MC simulations in
which the relevant parameters were varied.  Other uncertainties
arising from MC modelling as well as lepton identification and
momentum measurements have a substantially smaller impact.

\subsection{Results}
The results of this search are obtained by comparing the top quark
pair invariant mass ($m_{t\bar{t}}$) distribution with background-only
and signal-plus-background hypotheses.  In practice, the search is
done in two steps: in a first step the data is compared to the
Standard Model prediction, i.e.~the null hypothesis, using the {\sc
  BumpHunter}~\cite{Choudalakis:2011qn} algorithm. Since no excess is
found, in a second step a limit is set using a Bayesian
approach~\cite{Bertram:2000br} on the maximum allowed cross-section
times branching ratio for new physics as a function of $m_{t\bar{t}}$.
In this limit-setting step, 40 GeV-wide bins are used, a value close
to the mass resolution and limiting bin-by-bin statistical
fluctuations.  A single bin contains all events with $m_{t\bar{t}} >$
2.96 TeV.

\begin{figure}[ht]
\centering
\subfigure[]
{
\includegraphics[width=0.47\textwidth]{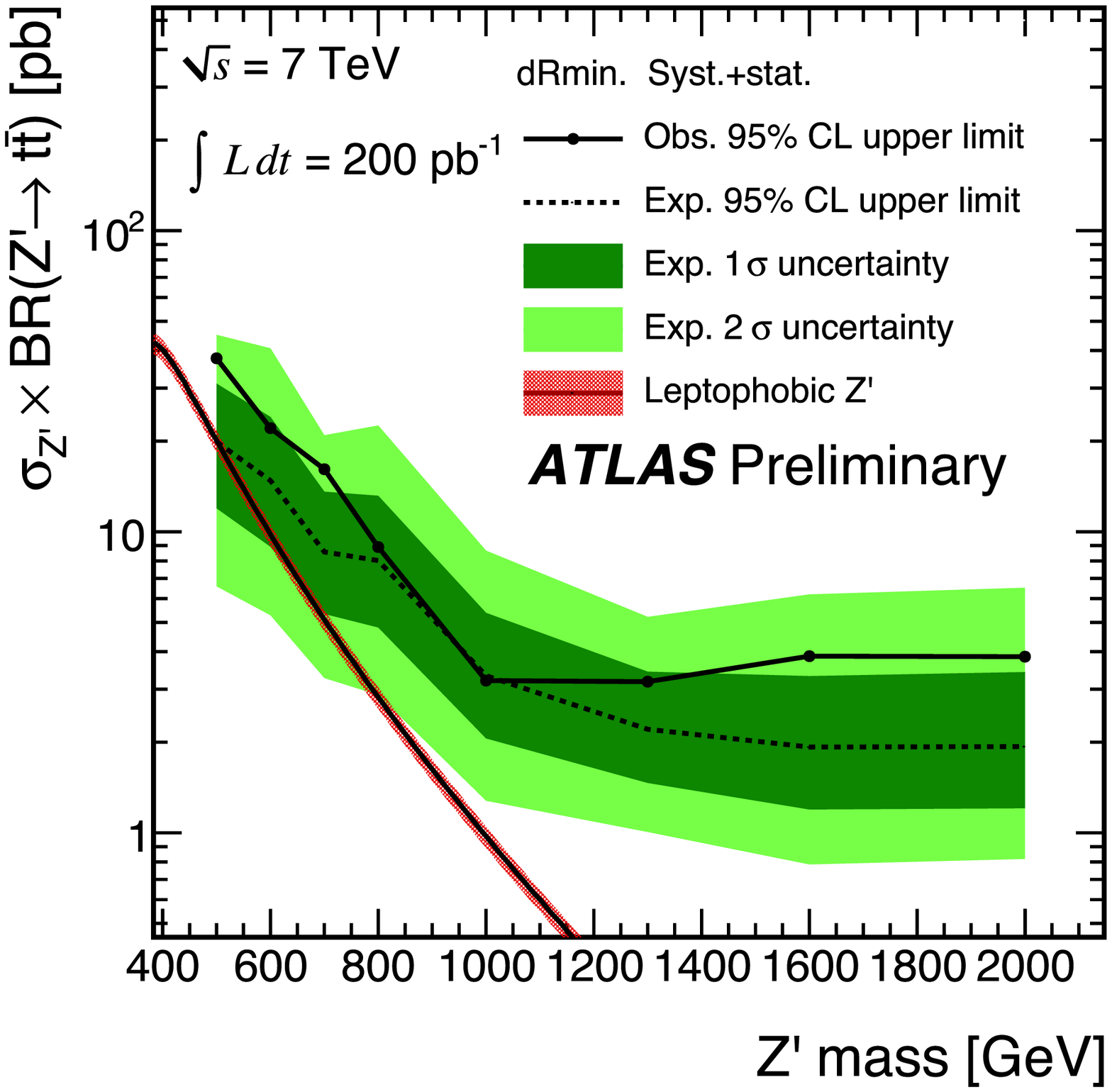}
}
\subfigure[]
{
\includegraphics[width=0.47\textwidth]{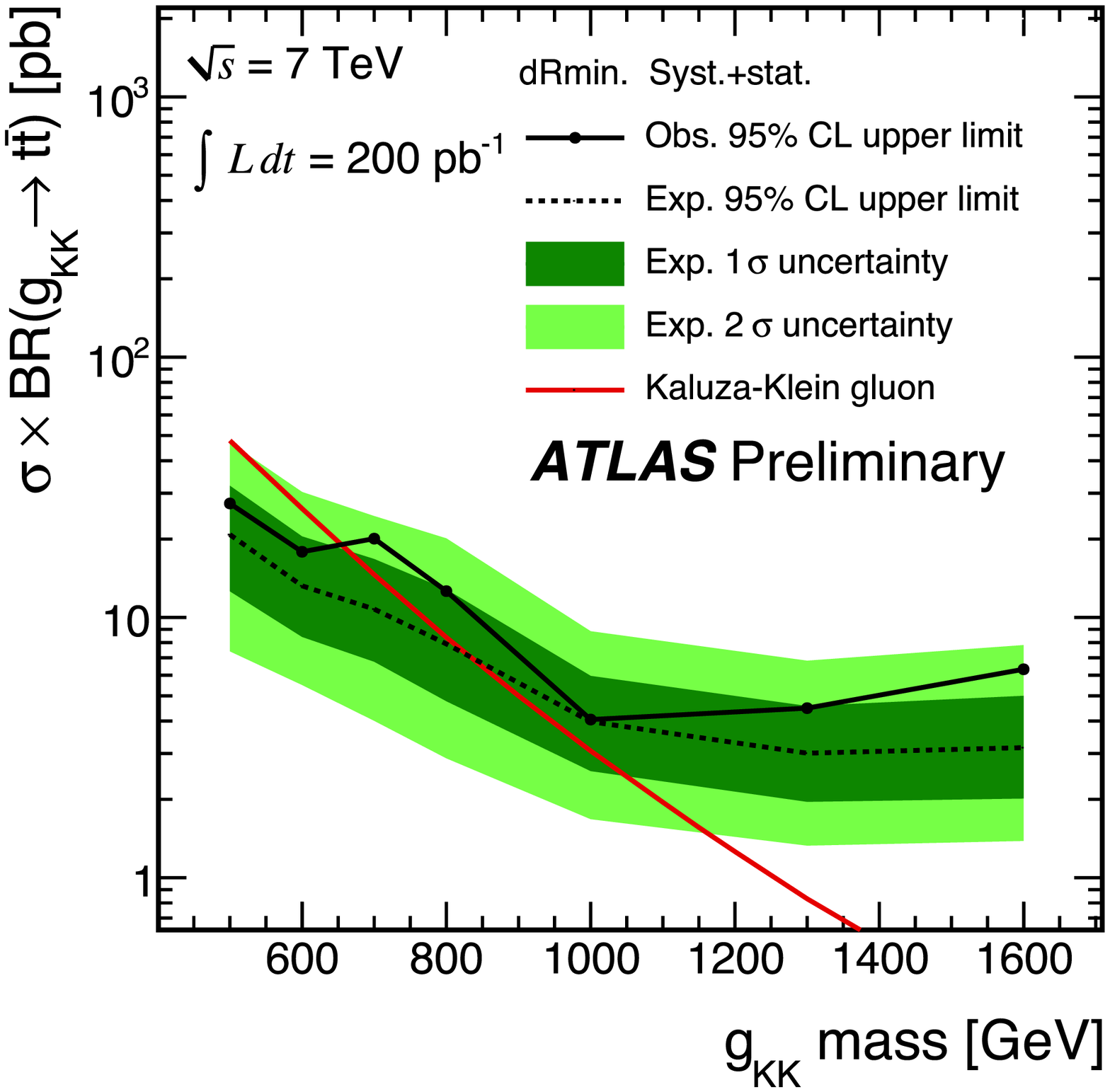}
}
\caption{Expected (dashed line) and observed (black points connected
  by a line) upper limits on $\sigma \times$~BR($Z' \to t \bar{t}$)
  (a) and $\sigma \times$~BR($g_{KK} \to t \bar{t}$) (b) using the
  $dRmin$ algorithm.  The dark and light green bands show the range in
  which the limit is expected to lie in 68\% and 95\% of experiments,
  respectively, and the red lines correspond to the predicted
  cross-section times branching ratio in the leptophobic topcolour and
  RS models.  The error bars on the topcolour cross-section curve
  represent the effect of the PDF uncertainty on the prediction.}
\label{fig:dr-sensitivity}
\end{figure}

The observed limits on narrow and wide resonances using the $dRmin$
mass reconstruction method are shown in Fig.~\ref{fig:dr-sensitivity}
together with the predicted cross-section times branching ratio for
the models considered and the expected sensitivity of the analysis.
The observed (expected) limit on $\sigma \times$~BR($Z' \to t
\bar{t}$) ranges from 38 (20) pb at $m_{Z'} = 500$ GeV to 3.2 (2.2) pb
at $m_{Z'} = 1300$ GeV.  While narrow resonances with production
cross-sections predicted by the leptophobic topcolour model cannot be
excluded, the analysis is already able to probe the few picobarn range
for masses close to 1 TeV.  The observed (expected) limit on $\sigma
\times$~BR($g_{KK} \to t \bar{t}$) ranges from 32 (24) pb at
$m_{g_{KK}} = 500$ GeV to 6.6 (2.9) pb at $m_{g_{KK}} = 1300$ GeV,
which excludes $g_{KK}$ resonances with mass below 650 GeV at 95\%
C.L.

\section{Summary and Conclusion}
Two searches for new phenomena involving top quarks are presented: a
search for a top partner in \ttbar\ events with large missing
transverse momentum, and a search for \ttbar\ resonances in 35~\ipb
and 200~\ipb of data, respectively. No evidence for a signal is
observed and 95\% C.L. limits are set on benchmark models. Both
analyses are currently being updated with more
luminosity~\cite{ttMETPaper}.

\end{document}